\documentclass[twocolumn,prl,aps,showpacs,superscriptaddress]{revtex4}
\usepackage{graphicx}
\usepackage{amssymb}

\begin{document}
\title{Coherent Backscattering with Nonlinear Atomic Scatterers}

\author{T. Wellens}
\affiliation{Institut Non Lin{\'e}aire de Nice, UMR 6618, 1361 route 
des Lucioles, F-06560 Valbonne}
\affiliation{Laboratoire Kastler Brossel, Universit{\'e} Pierre et Marie Curie,
4 Place Jussieu, F-75005 Paris}
\author{B. Gr{\'e}maud}
\author{D. Delande}
\affiliation{Laboratoire Kastler Brossel, Universit{\'e} Pierre et Marie Curie,
4 Place Jussieu, F-75005 Paris}
\author{C. Miniatura}
\affiliation{Institut Non Lin{\'e}aire de Nice, UMR 6618, 1361 route 
des Lucioles, F-06560 Valbonne}

\date{\today}

\begin{abstract}
We study coherent backscattering 
of a quasi-monochromatic laser
by a dilute gas of cold two-level atoms. We consider the perturbative regime
of weak intensities, where nonlinear effects arise from
{\em inelastic} two-photon scattering processes. Here, coherent backscattering
can be formed by interference between {\em three} different scattering
amplitudes. Consequently, if elastically 
scattered photons are filtered out from
the photodetection signal by means of suitable frequency-selective detection, we
find the nonlinear backscattering enhancement factor
to exceed the linear barrier two.
\end{abstract}

\pacs{42.25.Dd, 42.65-k, 32.80-t}

\maketitle

Light transport inside a nonlinear medium  gives rise to a wide
variety of phenomena, like pattern formation, four waves mixing,
self focusing effects, dynamical instabilities...\cite{boyd}. 
These effects are well described and
understood with the help of an intensity dependent susceptibility
(e.g. $\chi^{(3)}$ nonlinearity). However, in these approaches,
one usually discards the fact that interference phenomena in
disordered systems may significantly alter wave transport
properties. From this point of view, the
most striking systems for which one must combine both nonlinear
and disordered descriptions are coherent random lasers
\cite{cao}. Even if in this case one would require an
active ({\em i.e.} amplifying) medium, the key point is the
understanding of the mutual effects between multiple interferences
and nonlinear scattering. In particular, in the case of a dilute
medium, beyond deviations from the linear regime for the
propagation parameters (mean free paths, refraction indices), one
may expect modifications of interferential corrections to
transport (weak and strong localization regimes). A paradigmatic
example is given by {\em coherent backscattering} (CBS) experiments,
where an enhancement of the average intensity scattered around the
direction opposite to the incident wave is observed \cite{cbs}. 
In the linear
scattering regime (where the intensity of the scattered wave is
proportional to the incident intensity), CBS
arises from interference between {\em two} reversed scattering
paths visiting the same scatterers, but in opposite order. In this
case, the CBS enhancement factor, defined as the signal
detected in exact backscattering direction divided by the diffuse
background, never exceeds two. This maximum value is reached if
each pair of interfering paths has the same amplitude, 
and if single scattering can be suppressed.

Concerning the nonlinear regime, previous studies have been restricted to the
case of linear scatterers embedded in a uniform nonlinear
medium \cite{agra91,heid95,bress01}. Here, it has been shown that 
the maximum enhancement factor
remains two. As we will show in this letter, however, the situation 
drastically changes in the presence of
nonlinear {\em scattering} (in contrast to nonlinear {\em propagation}).
In particular, in the perturbative regime of at most one
scattering event with $\chi^{(3)}$ nonlinearity, CBS arises from
interference between {\em three} amplitudes. Depending on the sign of the nonlinearity,
this leads to an increase or decrease of the nonlinear
CBS enhancement factor compared to the linear value two.
The clearest manifestation of this effect, however, can be observed
if the linear component is filtered out from the backscattering
signal, such that
the latter exclusively originates from nonlinear 
processes. Then, interference between three amplitudes allows
values of the CBS enhancement factor up to three.

\begin{figure}
\centerline{\includegraphics[width=6cm]{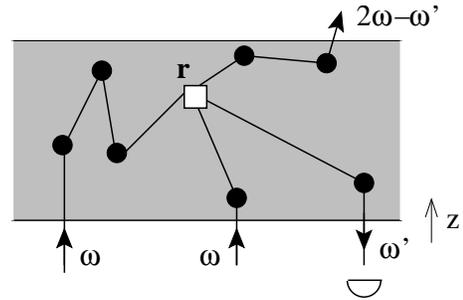}}
\caption{In the perturbative approach, we assume a single nonlinear 
two-photon scattering event ($\Box$), but arbitrarily many linear scattering
events ($\bullet$). One of the two photons is finally annihilated by the detector,
thereby defining the photodetection signal, whereas the other
one is scattered into an arbitrary direction.\label{nonl1}}
\end{figure}

In the case of a disordered medium consisting of
cold two-level atoms, the
linear and nonlinear component
of the backscattered light can be distinguished in terms of its frequency.
This is due to the fact that scattering is purely elastic in the
linear regime (described by single-photon scattering),
whereas nonlinear multiphoton processes in general change
the frequencies of the individual photons, giving rise to inelastic 
scattering \cite{cct}. 
In this letter, we will restrict ourselves to the perturbative regime of 
weak laser intensities which is described by scattering of two photons
\cite{wellens}.
A typical scattering path is represented in Fig.~\ref{nonl1}. 
Here, the two incoming photons propagate 
at first independently from each other to position $\bf r$ inside the disordered
atomic medium, where they undergo a nonlinear scattering 
event. One of the two outgoing photons then propagates back to the detector.
The possibility that the two photons  meet again at another atom can be neglected 
in the case of a dilute medium, similar to recurrent scattering
in the linear case \cite{recurr}. We can hence restrict our analysis
to processes like the one shown
in Fig.~\ref{nonl1}, with arbitrary number of linear scattering events before and 
after the nonlinear one. As further discussed below,
this perturbative approach is valid for small laser intensity and not too large 
optical thickness.

\begin{figure}
\centerline{\includegraphics[width=7.5cm]{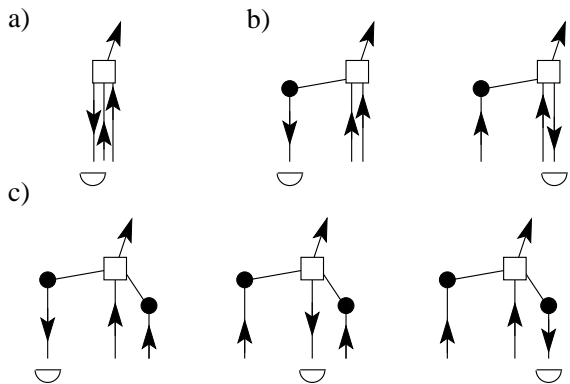}}
\caption{In the presence of nonlinear scattering ($\Box$),
there may be either (b) two, or (c) three
interfering amplitudes contributing to enhanced backscattering,
apart from single scattering (a), which only contributes to the background.
In general,
the case (c), which corresponds to maximum enhancement factor three,
is realized if either both incoming photons, or one incoming and the 
outgoing detected photon exhibit at least
one linear scattering event ($\bullet$) besides the nonlinear one.
\label{fig1}}
\end{figure}

To obtain the average intensity measured by the detector placed in backscattering
direction,
we must identify those scattering paths
whose interference survives the ensemble
average over the random positions of the scatterers. 
Again, we consider the case of a dilute medium, where the typical distance 
between two scattering events
(given by the linear mean free path $\ell$) is much larger than the 
wavelength $\lambda$.
First, we do not obtain interference between diagrams where the nonlinear
scattering event occurs at different atoms \cite{wellens}. 
Just as in the linear case, we {\em do} find interference
between diagrams where the path of the detected photon is reversed. 
This gives rise to an enhancement of the average detection signal 
in exact backscattering direction, on top of the diffuse background. 
In contrast to the linear
case, however, there may be up to {\em three} different interfering amplitudes, as
evident from the examples shown in
Fig.~\ref{fig1}.
In general, the
three-amplitudes case is realized if
both incoming photons, or
one incoming and the outgoing detected photon exhibit at least one linear 
scattering event besides the nonlinear one. 

The perturbative calculation of the backscattering signal relies on the fact that
the scattering amplitude of a multiple scattering process like the one shown in 
Fig.~\ref{nonl1} is obtained as a product of the scattering amplitudes for each
individual scattering event. The single-photon scattering events can be treated
by standard methods known from the theory of multiple scattering in the linear regime
\cite{multiple}, which we shortly summarize 
as follows. Here, the central quantity is the
average intensity $I_\omega({\bf r})$ inside the medium. 
It can be interpreted as 
the probability that an incoming photon with frequency $\omega$
reaches position $\bf r$ after
arbitrarily many linear scattering events, and fulfills the following version of 
the radiative transfer equation:
\begin{equation}
I_\omega({\bf r})=e^{-z/\ell}+{\mathcal{N}}|S_\omega|^2
\int_V d{\bf r'} |G_\omega({\bf r},{\bf r'})|^2 I_\omega({\bf r'}).\label{radtransf}
\end{equation}
Here, $z$ denotes the distance from the boundary of the medium
to $\bf r$ along the direction of the incident beam, 
and $\mathcal N$ the density of atoms which are
randomly distributed in the volume $V$.
Furthermore, 
\begin{equation}
S_\omega=\frac{i}{k(1-2i\delta/\Gamma)}\label{scatt}
\end{equation}
represents the scattering amplitude for
a photon with frequency $\omega=k c$ by a single atom,
where $\Gamma$ denotes the width of the atomic resonance at $\omega_{\rm at}$.
Eq.~(\ref{scatt}) is valid in the near-resonant case of small detuning 
$\delta=\omega-\omega_{\rm at}\ll\omega$.
For simplicity, we work with scalar photons, {\em i.e.}, we neglect the 
vectorial nature of the light field.
In this case, scattering is isotropic, with the total cross section given by
$\sigma=4\pi|S_\omega|^2$. This is not a crucial assumption - the following
treatment can be generalized to the vectorial case without serious 
difficulties. Finally, 
\begin{equation}
G_\omega({\bf r},{\bf r'})=\frac{e^{i n_\omega k|{\bf r}-{\bf r'}|}}
{|{\bf r}-{\bf r'}|}
\end{equation}
describes average propagation
between $\bf r$ and $\bf r'$ in the 
atomic medium, with refractive index given by 
\begin{equation}
n_\omega=1-\frac{\delta}{\Gamma k\ell}+\frac{i}{2k\ell}.\label{index}
\end{equation}
The imaginary part 
of $n_\omega$ is determined by
the linear mean free path $\ell=1/(\mathcal{N}\sigma)$ at frequency $\omega$.

The first term in Eq.~(\ref{radtransf}) 
represents the exponential attenuation of the coherent laser mode, {\em i.e.}, 
light which has penetrated to position
$\bf r$ without being scattered (Beer-Lambert law). 
The remaining term describes the diffuse intensity, {\em i.e.}, light which has 
been scattered at least once before reaching $\bf r$. 
The required solution of Eq.~(\ref{radtransf}) is obtained numerically by
iteration, starting from $I_\omega({\bf r})=0$.
In terms of the average intensity $I_\omega({\bf r})$,
the linear components of the backscattering signal
are readily obtained by integration over the medium \cite{multiple}.

The second ingredient needed for our nonlinear perturbative analysis is scattering
of two photons by a single atom \cite{wellens}. Here, the scattered light
contains an elastic component, with the same frequency $\omega$ as the incident light,
and an inelastic component with power spectrum
\begin{equation}
P(\omega') = \frac{\Gamma}{4\pi}
\left|\frac{1}{\delta'+i\Gamma/2}+
\frac{1}{2\delta-\delta'+i\Gamma/2}\right|^2\label{specin},
\end{equation}
where $\delta'=\omega'-\omega_{\rm at}$ denotes the final detuning.
The ratio between the inelastic and elastic
component is given by the saturation parameter $s=2\Omega^2/(4\delta^2+\Gamma^2)$
(with the Rabi frequency $\Omega$), which measures the 
intensity of the incident laser in units of the saturation intensity \cite{cct}.

Combining scattering of two photons by a single atom 
with the technique for treating the linear case outlined above, we arrive at the
following expression for the inelastic background component at frequency 
$\omega'$ (postponing a more detailed derivation to a further publication):
\begin{equation}
L_{\rm in}(\omega')= 
s P(\omega')\int_V \frac{d{\bf r}}{A\ell}
\left(2I_\omega^2({\bf r})-e^{-2z/\ell}\right) I_{\omega'}({\bf r}),\label{nonlinl2}
\end{equation}
where $A$ denotes the transverse area of the medium.
The term $2I_\omega^2({\bf r})-e^{-2z/\ell}$ in Eq.~(\ref{nonlinl2})
can be interpreted as the average {\em squared} intensity at position $\bf r$ 
inside the slab,
which is larger than the  square $I_\omega^2({\bf r})$ of the average intensity due to
speckle fluctuations \cite{fluc}. The squared intensity induces 
a nonlinear atomic response, emitting a photon with frequency distribution 
$P(\omega')$. Due to time reversal symmetry, the diffusion of this photon from
$\bf r$ to the detector is given by the same expression
$I_{\omega'}({\bf r})$ which describes diffusion of incoming photons to 
$\bf r$. 
Note that, in the perturbative regime, the inelastic component contains
nonlinear scattering ({\em i.e.} nonlinearity of the scattering cross section),
but not nonlinear average propagation ({\em i.e.} nonlinearity of the
mean free path). Since the latter leaves the photon 
frequency unchanged, it only affects the elastic component, 
which we assume to be filtered out from the detection signal.
 
Concerning the calculation of the interference contribution
giving rise to enhanced backscattering,
the linear average intensity 
$I_\omega({\bf r})$, Eq.~(\ref{radtransf}),
must be generalized in order to account for
different frequencies $\omega$ and $\omega'$ in the interfering paths:
\begin{eqnarray}
g_{\omega,\omega'}({\bf r}) & = & e^{i(n_\omega-n_{\omega'}^*)kz}+{\mathcal N}
S_\omega S_{\omega'}^*\nonumber\\
& & \times\int_V d{\bf r'}G_\omega({\bf r},{\bf r'})G_{\omega'}^*({\bf r},{\bf r'})
g_{\omega,\omega'}({\bf r'}).
\label{radtransfc}
\end{eqnarray}
This function describes the ensemble-averaged product of two probability 
amplitudes, 
one representing an incoming photon with frequency $\omega$ 
propagating to position $\bf r$, and the other one the complex conjugate
of a photon with frequency $\omega'$ propagating
from $\bf r$ to the detector. 
If $\omega\neq\omega'$, a nonvanishing phase difference 
between these amplitudes arises due to both scattering and 
average propagation in the medium,
since both the complex scattering amplitude, Eq.~(\ref{scatt}),
and the refractive index, Eq.~(\ref{index}), depend on frequency.
In contrast, the phase difference due to free propagation ({\em i.e.} in the vacuum) 
can be neglected if $\Gamma\ell\ll c$, which is fulfilled for typical experimental 
parameters \cite{thierry}. In the case $\omega=\omega'$, 
$g_{\omega,\omega}({\bf r})=I_\omega({\bf r})$ reduces to the average
intensity, see Eq.~(\ref{radtransf}).
In terms of the iterative solution of Eq.~(\ref{radtransfc}), we find
the following expression for the interference term:
\begin{eqnarray}
C_{\rm in}(\omega') & = & 4 s P(\omega')\int_V \frac{d{\bf r}}{A\ell}
\Bigl[I_\omega({\bf r})
|g_{\omega,\omega'}({\bf r})|^2\Bigr.\nonumber\\
& & -e^{-z/\ell} {\rm Re}\left\{e^{i(n_\omega-n_{\omega'}^*)kz}
g_{\omega,\omega'}^*({\bf r})\right\}\nonumber\\
& & \Bigl.-\left(I_\omega({\bf r})-e^{-z/\ell}\right)e^{-z/\ell-z/\ell'}\Bigr],
\label{nonlinc2}
\end{eqnarray}
with $\ell'$ the linear mean free path at frequency $\omega'$.
In order to verify that the interference of either three or two amplitudes is correctly
taken into account in Eq.~(\ref{nonlinc2}), it is useful to write the average
intensity $I=\exp(-\zeta)+I_D$ as a sum of the coherent mode $\exp(-\zeta)$ 
(with $\zeta=z/\ell$) plus diffuse intensity $I_D$. In the case $\omega'=\omega$,
we obtain from Eqs.~(\ref{nonlinl2},\ref{nonlinc2}):
\begin{eqnarray}
L_{\rm in} & \propto & \bigl<e^{-3\zeta}+5I_De^{-2\zeta}
+~6I_D^2e^{-\zeta}+2I_D^3\bigr>,\nonumber\\
C_{\rm in} & \propto &\bigl<
\underbrace{\phantom{e^{-3\zeta}+\!}}_{\displaystyle (a)}~\underbrace{4I_De^{-2\zeta}}_{\displaystyle (b)}+\underbrace{12I_D^2e^{-\zeta}+4I_D^3}_{\displaystyle (c)}\bigr>,
\label{laddcross}
\end{eqnarray} 
where the brackets denote the integral over the volume $V$ of the medium, and
(a,b,c) correspond to the three cases
shown in Fig.~\ref{fig1}, identified by different powers of diffuse or 
coherent light. As expected, the three-amplitudes case (c) implies an 
interference term twice as large as the background, whereas the case (b) leads to the
factor $4/5$ known from the scalar two-atom solution (since one of the two
interfering amplitudes is twice as large as the other one \cite{wellens}).
Finally, as it should be, the single scattering term is absent in the interference term
$C_{\rm in}$.

\begin{figure}
\centerline{\includegraphics[height=7.5cm,angle=270]{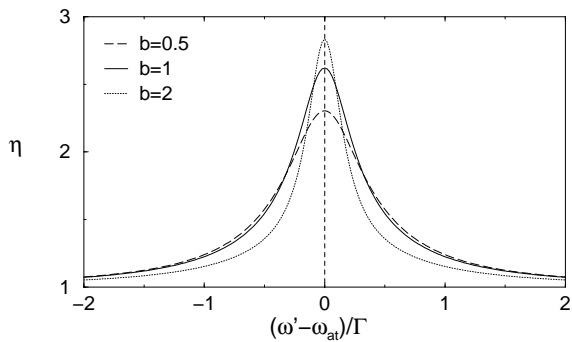}}
\caption{Spectral dependence of the enhancement factor for coherent backscattering
from a slab with optical thickness
$b=0.5$ (dashed line), $b=1$ (solid line), and $b=2$ (dotted line) in the
perturbative regime of small saturation parameter $s$. The initial
frequency $\omega=\omega_{\rm at}$ is chosen on resonance ({\em i.e.} $\delta=0$).
The vertical line displays the frequency of elastically scattered photons,
which must be filtered out in order to observe
an enhancement factor larger than two.
\label{spek}}
\end{figure}

Using Eqs.~(\ref{nonlinl2},\ref{nonlinc2}), we now calculate the 
CBS enhancement factor $\eta$ for the
frequency component $\omega'$:
\begin{equation}
\eta(\omega')  = 1+\frac{C_{\rm in}(\omega')}{L_{\rm in}(\omega')}.
\end{equation}
The result is shown in Fig.~\ref{spek}, for a slab geometry with 
three different values of the optical thickness,
$b=0.5$, $1$, and $2$.
Evidently, the largest values of the
enhancement factor are obtained if the
final frequency $\omega'$ approaches the initial one $\omega$, 
since then the dephasing due
to different frequencies vanishes. 
In this case, the enhancement factor is completely
determined by the relative weights between the one-, two- and three-amplitudes
cases shown in Fig.~\ref{fig1}, cf. Eq.~(\ref{laddcross}). 
As evident from the dashed line in Fig.~\ref{spek},
already at the rather moderate value $b=0.5$ of the optical thickness, the
three-amplitudes case is sufficiently strong in order to increase the maximum
enhancement factor above the linear barrier $\eta=2$. With increasing optical thickness
(and, if necessary, decreasing saturation parameter, in order to stay in the domain of
validity of the perturbative approach, see below), the
number of linear scattering events increases,
which implies that the three-amplitudes case increasingly dominates, 
see Fig.~\ref{fig1}. In this limit, the enhancement factor
approaches the maximum value three. At the same time, however, a larger number of 
scattering events
also leads to stronger dephasing due to
different frequencies, $\omega'\neq\omega$. This results in a narrower shape of
$\eta$ as a function of $\omega'$ for larger optical thickness.
Nevertheless, as evident from Fig.~\ref{spek},
the enhancement factor remains larger than two in a significant range of frequencies
$\omega'$. 
Hence, an experimental demonstration of three-amplitudes 
interference requires a sufficiently narrow spectral filter, which should be placed as
close to the laser frequency $\omega$ as possible, but far enough to filter out the
elastic component. We have checked that, without filtering,
the enhancement factor {\em decreases} as a function of $s$
(due to the negative sign of the elastic nonlinear component), in qualitative
agreement with the experiment \cite{thierry}. For a quantitative comparison,
the elastic component, the geometry of the medium, and the polarization
of the photons
must be taken into account, which will be presented 
elsewhere.

Finally, the domain of validity of our
perturbative approach, assuming a single nonlinear scattering event,
remains to be discussed. A rough quantitative estimation can be
given as follows:
if $p_1$ (resp. $p_{2+}$) denotes the probability
for a backscattered photon to undergo one (resp. more than two) 
nonlinear scattering event,
the perturbative condition reads $p_{2+}\ll p_1$.
If we assume for all scattering events the same probability
$s$ to be nonlinear (thereby neglecting the
inhomogeneity of the local intensity), we obtain 
$p_1\simeq \langle N\rangle s$, and $p_{2+}\simeq \langle N^2\rangle s^2$,
where $N$ denotes the total number of scattering events, and
$\langle\dots\rangle$ the statistical average over all backscattering
paths. Evidently, $N$ and $N^2$ are expected to increase when increasing
the optical thickness $b$. For a slab geometry, we have
found numerically that 
$\langle N\rangle\propto b$ and $\langle N^2\rangle\propto b^3$ 
(in the limit of large $b$), concluding that the perturbative
treatment is valid if $sb^2\ll 1$. Let us note that
we remain well below the threshold $sb^2\simeq 1$ at which speckle
fluctuations in a nonlinear medium become unstable \cite{skipetrov}.

A natural way how to extend this work is to give up the perturbative
assumption, and admit more than one nonlinear scattering event.
This is necessary in order to describe media with large optical thickness,
even at small saturation. Since 
the number of interfering amplitudes is expected to increase
if more than two photons are connected by nonlinear scattering events,
it may be possible to observe even larger
enhancement factors - especially in the case of amplifying scatterers where the
nonlinearity appears with positive sign. Finally, 
the effect demonstrated in this letter might also be relevant for
strong localization in the nonlinear regime.

It is a pleasure to thank Cord M\"uller for fruitful discussions.
T.W. has been supported by the DFG Emmy Noether program.
Laboratoire Kastler Brossel is laboratoire de l'Universit\'e Pierre et Marie
Curie et de l'Ecole Normale Superieure, UMR 8552 du CNRS.

\end{document}